\documentclass[12pt]{article}
\usepackage{amssymb}
\usepackage{amsthm}
\usepackage{eucal}
\theoremstyle{plain}
\newtheorem{theorem}{Theorem}[section]
\newtheorem{definition}{Definition}[section]
\newtheorem{lemma}[theorem]{Lemma}
\newtheorem{corollary}[theorem]{Corollary}

\newtheorem*{m-p}{Rule of Inference}

\begin{document}
\voffset=-20pt
\hoffset=-60pt
\textwidth=180mm
\textheight=210mm

\markright{\it International Journal of Theoretical Physics\/\rm,
\bf 42\sl, No.~12 \rm (2003)}

\vbox to 2cm{\vfill}

\begin{flushleft}
{\LARGE\bf Quantum Implication Algebras}
\end{flushleft}

\bigskip
\bigskip
\bigskip
\parindent=2cm\hangindent=4cm\baselineskip=15pt

\bf Norman D.~Megill$\,$\footnote{Boston Information Group,
Belmont MA 02478, U.~S.~A.; E-mail: nm@alum.mit.edu;
Web page: http://www.metamath.org}
and
Mladen Pavi\v ci\'c$\,$\footnote{University of Zagreb,
Gradjevinski fakultet, Ka\v ci\'ceva 26, HR-10001 Zagreb, Croatia;\break
E-mail: mpavicic@irb.hr; Web page: http://m3k.grad.hr/pavicic}

\rm
\medskip
\vrule height .3ex width 11.70493cm depth -.05ex

\parindent=2cm\hangindent=2cm\baselineskip=15pt\medskip
Quantum implication algebras without complementation are formulated
with the same axioms for all five quantum implications.
Previous formulations of orthoimplication, orthomodular implication,
and quasi-implication algebras are analysed and put in perspective to
each other and our results.

\medskip

\vrule height .3ex width 11.71043cm depth -.05ex

\smallskip


\smallskip


\parindent=4.1cm
semi-orthomodular lattices, orthomodular lattices, 

\parindent=4.1cm
orthomodular implication algebras, join semilattices.

\parindent=2cm
\smallskip

\vrule height .3ex width 11.71043cm depth -.05ex

\vfill\eject
\markright{\rm N.~Megill and M.~Pavi\v ci\'c \
     \it Quantum Implication Algebras}

\parindent=20pt\hangindent=0pt
\vbox to 3mm{\vfill}
\baselineskip=17pt
\section{\large Introduction}
\label{sec:intro}

It is well-known that there are five operations of implication
in an orthomodular lattice which all reduce to the classical
implication in a distributive lattice. \cite{kalmb83}
It was therefore believed that implication algebras for these
implications must all be different and such different algebras
have explicitly been defined in the literature.
\cite{clark73,piziak74,abbott76,gaca80,harde81a,langer1}

In a previous paper \cite{mpijtp98} we have shown that
one can formulate quantum implication algebras with ``negation''
[(ortho)com\-ple\-mentation] with the same axioms for all five quantum
implications. We arrived at such a formulation of implication
algebras by using a novel possibility, given in
Refs.~\cite{mpqo01} and \cite{mpqo02}, of
defining different quantum operations by each other.
Implicitly, the latter possibility provides us a direct way of
formulating quantum algebras without complementation and in this
paper we give it.

To do so, we were prompted by a recent formulation of an
implication algebra. \cite{langer1} The authors formulate an
algebra based on the Dishkant implication previously considered
by \cite{kimble69,abbott76,gaca80} and cited by \cite{harde81a,mpijtp98}.
There are also other quantum implication algebras given by
\cite{finch70,clark73,piziak74,harde81a,harde81b,gaca80,mpijtp98}
and others. In this paper we show how are all these algebras
interrelated.

\bigskip

\section{\large Preliminaries}
\label{sec:latt}

Let us first repeat a definition of an orthomodular
lattice. \cite{mpqo02}

\begin{definition}\label{D:latt} An orthomodular lattice ({\rm OML}) is
an algebraic structure $\langle L,\cup,^{\perp}\rangle$ in which
the following conditions are satisfied for any $a,b,c\in L$:

\smallskip
{\bf L1.}\quad $a\le a^{\perp\perp}\quad \&\quad a^{\perp\perp}\le a$

\smallskip
{\bf L2.}\quad $a\le a\cup b \quad \&\quad b\le a\cup b$

\smallskip
{\bf L3.}\quad $a\le b\quad \&\quad b\le a\quad\Rightarrow\quad a=b$

\smallskip
{\bf L4.}\quad $a\le 1$

\smallskip
{\bf L5.}\quad $a\le b\quad\Rightarrow\quad b^{\perp}\le a^{\perp}$

\smallskip
{\bf L6.}\quad$a\le b\quad \&\quad b\le c\quad\Rightarrow\quad a\le
c$

\smallskip
{\bf L7.}\quad$a\le c\quad \&\quad b\le c\quad\Rightarrow\quad a\cup
b\le c$

\smallskip
{\bf L8.}\quad $a\rightarrow_ib\>=\>1$$\qquad\Rightarrow
\qquad$$a\le b\qquad\qquad(i=1,\!...,5)$

\parindent=0pt
\smallskip
where $a\le b{\buildrel\rm def\over\Leftrightarrow}a\cup b=b$,
$1\ {\buildrel\rm def\over=}\ a\cup a^{\perp}$.
Also
$$a\cap b\ {\buildrel\rm def\over=}\
(a^{\perp}\cup b^{\perp})^{\perp},
\qquad 0\ {\buildrel\rm def\over=}\ a\cap a^{\perp}.$$
\smallskip\parindent=0pt
and the implications $a\rightarrow_ib\ (i=1,\dots,5)$ are defined
as follows

\rm\parindent=20pt

\medskip
$a\rightarrow _1 b \ \ {\buildrel\rm def\over=} \ \
a^\perp \cup(a\cap b)$ \hfill (Sasaki)

\medskip
$a\rightarrow _2 b \ \ {\buildrel\rm def\over=} \ \
b\cup(a^\perp \cap b^\perp )$  \hfill  (Dishkant)

\medskip
$a\rightarrow _3 b \ \ {\buildrel\rm def\over=} \ \
((a^\perp \cap b)\cup(a^\perp \cap b^\perp ))\cup\bigl(a\cap(a^\perp \cup b)
\bigr)$  \hfill  (Kalmbach)

\medskip
$a\rightarrow _4 b \ \ {\buildrel\rm def\over=} \ \
((a\cap b)\cup(a^\perp \cap b))\cup\bigl((a^\perp \cup b)\cap
b^\perp \bigr)$   \hfill  (non-tollens)

\medskip
$a\rightarrow _5 b\ \ {\buildrel\rm def\over=} \ \
((a\cap b)\cup(a^\perp \cap b))\cup(a^\perp \cap b^\perp )$ \hfill
(relevance)
\end{definition}

The following theorem is well-known.

\medskip
\begin{theorem}\label{th:od}
The equation $a^\perp=a\rightarrow_i 0$
is true in all orthomodular lattices for $i=1,\dots,5$.
\end{theorem}

\begin{proof}The proof is straightforward and we omit it.
\end{proof}

There are 6 Boolean-equivalent expressions for implication in an OML.
In addition to the 5 {\em quantum} implications above, which are
distinguished by satisfying L8 (also known as the
Birkhoff-von Neumann requirement), we have
the {\em classical} implication that does not satisfy L8 in
every OML:

\medskip
$a\rightarrow _0 b \ \ {\buildrel\rm def\over=} \ \
a^\perp \cup b$ \hfill (classical)

\section{\large Implication algebras based on the Dishkant
implication}\label{sec:oia}

Two kinds of implicational algebras based on the Dishkant implication
$\rightarrow _2$ have been proposed in the literature:  {\em
orthoimplication algebras} \cite{abbott76} and {\em orthomodular
implication algebras} \cite{langer1}.  In this section we
summarise the two systems and some of their principle results, which are
proved in their respective articles.  As much as is practical we attempt
to use the terminology of the authors of those articles.

\begin{definition}\label{D:oia} \cite{abbott76} An orthoimplication
algebra ({\rm OIA}) is an algebraic structure $\langle{\mathcal A},\cdot
\rangle$ with a single binary operation that satisfies:

\smallskip
{\bf OI1}\quad$(ab)a=a$

\smallskip
{\bf OI2}\quad$(ab)b=(ba)a$

\smallskip
{\bf OI3}\quad$a((ba)c)=ac$

\end{definition}

\begin{definition}\label{D:omia} \cite{langer1} An orthomodular
implication algebra ({\rm OMIA}) is an algebraic structure $\langle{\mathcal
A},\cdot,1 \rangle$ with binary operation $\cdot$ and constant $1$
that satisfy:

\smallskip
{\bf O1}\quad$aa = 1$

\smallskip
{\bf O2}\quad$a(ba) = 1$

\smallskip
{\bf O3}\quad$(ab)a = a$

\smallskip
{\bf O4}\quad$(ab)b = (ba)a$

\smallskip
{\bf O5}\quad$(((ab)b)c)(ac) = 1$

\smallskip
{\bf O6}\quad$(((((((((ab)b)c)c)c)a)a)c)a)a = (((ab)b)c)c$

\end{definition}

We note that the theorem $aa=bb$ holds in both systems, and it can be
proved under OMIA without invoking axiom O1.  Thus we may treat the
constant $1$ of OMIA as a defined term $1\>=^{\rm def}\>aa$ (making
axiom O1 redundant), or we may extend OIA with a constant $1$ (and add
an axiom $aa=1$ for it).  For ease of comparing the two systems, we
choose the first approach and henceforth shall consider $1$ to be a
defined term in OMIA.

Both OIA and OMIA are {\em sound} for the Dishkant implication in the
sense that if the binary operation $\cdot$ is replaced throughout by
$\rightarrow _2$, each axiom becomes an equation that holds in all OMLs.
Thus each of these systems corresponds to a (not necessarily complete)
Dishkant implicational fragment of OML theory.

A {\em join semilattice} is a partially-ordered set that is bounded
above and in which every pair of elements has a least upper bound.  Both
OIA and OMIA induce join semilattices $\langle{\mathcal A},\cup,1
\rangle$ under the definitions $a\cup b\>=^{\rm def}\>(ab)b$ and
$1\>=^{\rm def}\>aa$, with the partial order defined by $a\le
b\>\Leftrightarrow^{\rm def}\>a\cup b=b\>\Leftrightarrow\>ab=1$.

The algebras OIA and OMIA also induce, respectively, more specialised
associated structures called {\em semi-orthomodular lattices} and {\em
orthomodular join semilattices}.  These are defined as follows.

\begin{definition}\label{D:ojs}
\cite{langer1} An orthomodular join semilattice ({\rm OJS}) is an
algebraic structure $\langle{\mathcal A},\cup,1,\langle {}^\perp_x; x
\in {\mathcal A} \rangle \rangle$ where $\langle{\mathcal A},\cup,1
\rangle$ is a join semilattice and $\langle {}^\perp_x; x \in {\mathcal
A} \rangle$ is a sequence of unary operations, one for each member $x$
of ${\mathcal A}$, such that the structure $\langle F_x, \cup,
{}^\perp_x\rangle$ is an orthomodular lattice, where $F_x \>=^{\rm
def}\> \{y|x\le y\}$ the principal filter of ${\mathcal A}$ generated by
$x$.
\end{definition}

\begin{definition}\label{D:sol} \cite{abbott76} A semi-orthomodular
lattice ({\rm SOL}) is an {\rm OJS} with the further requirement

\smallskip
{\bf C}\qquad$a\le b\le c \quad\Rightarrow\quad c^\perp_b=c^\perp_a\cup b$.

\end{definition}

\begin{theorem}\label{th:oia-sol}
\cite{abbott76} (i) Every {\rm OIA} induces an {\rm SOL} under the definition
$a^\perp_b \>$ $=^{\rm def}\> ab$ for $a \in F_b$.  (ii) Every {\rm SOL} induces
an {\rm OIA} under the definition $ab \>=^{\rm def}\> (a \cup b)^\perp_b$.
\end{theorem}

\begin{theorem}\label{th:omia-ojs}
\cite{langer1} (i) Every {\rm OMIA} induces an {\rm OJS} under the definition
$a^\perp_b \>=^{\rm def}\> ab$ for $a \in F_b$.  (ii) Every {\rm OJS} induces
an {\rm OMIA} under the definition $ab \>=^{\rm def}\> (a \cup b)^\perp_b$.
\end{theorem}

\section{\large Relationship between algebras OIA and OMIA}
\label{sec:rel}

In this section we show that the axioms of OMIA can be derived
from the axioms of OIA but not vice-versa.

\begin{theorem}\label{th:oia-omia}Every {\rm OIA} is an {\rm OMIA}.
\end{theorem}
\begin{proof}
To show this, we derive the axioms of OMIA from the axioms of OIA.

O1 is Lemma 1(i) of \cite{abbott76}.

O2 is Lemma 1(v) of \cite{abbott76}.

O3 is the same as OI1.

O4 is the same as OI2.

O5 can be expressed as $(a\cup b)c \le ac$.  From Th. 2 of
\cite{abbott76}, $a \le a\cup b$.  Therefore from Th.~1 of
\cite{abbott76}, $(a\cup b)c \le ac$.

We can now assume that Lemma~4 of \cite{langer1}, which makes use of
O1---O5 only, holds in OIA.

The associative law $a\cup (b\cup c)=(a\cup b)\cup c$ is derived as
follows.  Relations OL1---OL5 of \cite{mpqo02} correspond to
(v)---(viii) and (x) of Lemma~4 of \cite{langer1}.  In \cite{mpqo02} the
associative law L2a is proved using OL1---OL5 only, so it also holds in
OIA.  The associative law allows us to omit parentheses and (with the
help of OI2) disregard the order of joins in what follows.

O6 can be expressed as $((((a\cup b\cup c)c)\cup a)c)\cup a=a\cup b\cup c$.
The OM4 part of
Th.~4 of \cite{abbott76} contains a proof of
\begin{eqnarray}
     x\le y \quad\&\quad y \le z & \Rightarrow & y\cup ((y\cup (zx))x)=z \nonumber
\end{eqnarray}
or using OI2 and rewriting,
\begin{eqnarray}
     x\le y \quad\&\quad y \le z & \Rightarrow & (((zx)\cup y)x)\cup y=z \nonumber
\end{eqnarray}
We substitute $c$ for $x$, $a\cup c$ for $y$,
and $a\cup b\cup c$ for $z$:
\begin{eqnarray}
     &c\le a\cup c \quad\& & a\cup c\le a\cup b\cup c \quad\Rightarrow \nonumber\\
     && ((((a\cup b\cup
           c)c)\cup a\cup c)c)\cup a\cup c=a\cup b\cup c \nonumber
\end{eqnarray}
The hypotheses are satisfied by Th.~2 of \cite{abbott76}, so we have
\begin{eqnarray}
     ((((a\cup b\cup c)c)\cup a\cup c)c)\cup a\cup c&=&a\cup b\cup c  \nonumber
\end{eqnarray}
  From (v), (viii), and (x) of Lemma~4 of \cite{langer1}
we have $a\le b \Rightarrow a\cup b=b$.
By Lemma 1(v) of \cite{abbott76},  $c\le xc$ so $(xc)\cup c=xc$.
Applying this twice, the above becomes
\begin{eqnarray}
     ((((a\cup b\cup c)c)\cup a)c)\cup a&=&a\cup b\cup c   \nonumber
\end{eqnarray}
which is O6.
\end{proof}

On the other hand, it turns out that not every {\rm OMIA} is an {\rm OIA}.

\begin{theorem}\label{th:omia-ne-oia} There exist {\rm OMIA}s that are not
{\rm OIA}s.
\end{theorem}
\begin{proof}
Table~\ref{tab:chcounterex}(i) specifies an OMIA, i.e.\ any assignment
to the variables in the OMIA axioms will result in an equality using the
operation values in this table.  On the other hand, this OMIA is not an
OIA.  To see this, choose $a=5$, $b=2$, and $c=0$ in Axiom OI3.  Then
$a((ba)c)=5((2\cdot 5)0)=5(3\cdot 0)=5\cdot 2=10$ but $ac=5 \cdot 0=4$.
\end{proof}

\begin{table}[hbtp]
{\begin{center}
\begin{tabular}{c|cccccccccccc}
\multicolumn{1}{r|}{\lower.5ex\hbox{$a \diagdown^{\textstyle\!\!b}$}
        $\!\!\!$} &
          0 & 1 & 2 & 3 & 4 & 5 & 6 & 7 & 8 & 9 & 10 & 11 \\
\hline
      0 &  {\bf 1} &  1 &  1 &  1 &  1 &  1 &  1 &  1 &  1 &  1 &  1 &  1 \\
      1 &  {\bf 0} &  {\bf 1} &  {\bf 2} &  {\bf 3} &  {\bf 4} &  {\bf 5} &
             {\bf 6} &  {\bf 7} &  {\bf 8} &  {\bf 9} & {\bf 10} & {\bf 11} \\
      2 &  {\bf 3} &  1 &  {\bf 1} &  3 &  1 &  3 &  1 &  3 &  1 &  3 &  1 &  3 \\
      3 &  {\bf 2} &  1 &  2 &  {\bf 1} &  4 &  {\bf 8} &  6 & {\bf 10} &  8 &
               {\bf 4} & 10 &  {\bf 6} \\
      4 &  {\bf 5} &  1 &  {\bf 6} &  3 &  {\bf 1} &  5 &  6 &  7 &  8 &  {\bf 3}
                 & 10 & 11 \\
      5 &  {\bf 4} &  1 & 10 &  1 &  4 &  {\bf 1} &  6 & 10 &  1 &  4 & 10 &  6 \\
      6 &  {\bf 7} &  1 &  {\bf 4} &  3 &  4 &  5 &  {\bf 1} &  7 &  8 &  9 & 10 &
              {\bf 3} \\
      7 &  {\bf 6} &  1 &  8 &  1 &  4 &  8 &  6 &  {\bf 1} &  8 &  4 &  1 &  6 \\
      8 &  {\bf 9} &  1 & {\bf 10} &  3 &  4 &  {\bf 3} &  6 &  7 &  {\bf 1} &  9
                & 10 & 11 \\
      9 &  {\bf 8} &  1 &  6 &  1 &  1 &  8 &  6 & 10 &  8 &  {\bf 1} & 10 &  6 \\
      10 & {\bf 11} &  1 &  {\bf 8} &  3 &  4 &  5 &  6 &  {\bf 3} &  8 &  9 &
               {\bf 1} & 11 \\
      11 & {\bf 10} &  1 &  4 &  1 &  4 &  8 &  1 & 10 &  8 &  4 & 10 &  {\bf 1} \\
\end{tabular}
\end{center}}
    \caption{(i) Example of an orthomodular implication algebra (OMIA),
    with operation $ab$, that is not an orthoimplication algebra (OIA).
    (ii) The bold entries specify the partial functions
     $a^\perp_b$ for the OJS of Figure~\ref{fig:chcounterex}.
\label{tab:chcounterex}}
\end{table}

\begin{figure}[htbp]\centering
    \setlength{\unitlength}{1pt}
    \begin{picture}(330,100)(0,0)

      \put(100,5){
        \begin{picture}(60,90)(0,0)
          \put(60,90){\line(-2,-1){60}}
          \put(60,90){\line(-1,-1){30}}
          \put(60,90){\line(0,-1){30}}
          \put(60,90){\line(1,-1){30}}
          \put(60,90){\line(2,-1){60}}
          \put(60,0){\line(-2,1){60}}
          \put(60,0){\line(-1,1){30}}
          \put(60,0){\line(0,1){30}}
          \put(60,0){\line(1,1){30}}
          \put(60,0){\line(2,1){60}}

          \put(0,60){\line(0,-1){30}}
          \put(0,60){\line(1,-1){30}}
          \put(0,60){\line(2,-1){60}}
          \put(0,60){\line(3,-1){90}}

          \put(30,60){\line(-1,-1){30}}
          \put(60,60){\line(-1,-1){30}}
          \put(90,60){\line(-1,-1){30}}
          \put(120,60){\line(-1,-1){30}}

          \put(120,30){\line(0,1){30}}
          \put(120,30){\line(-1,1){30}}
          \put(120,30){\line(-2,1){60}}
          \put(120,30){\line(-3,1){90}}

          \put(60,-5){\makebox(0,0)[t]{$0$}}
          \put(-5,30){\makebox(0,0)[r]{$9$}}
          \put(38,30){\makebox(0,0)[l]{$11$}}
          \put(65,28){\makebox(0,0)[l]{$5$}}
          \put(96,30){\makebox(0,0)[l]{$7$}}
          \put(125,30){\makebox(0,0)[l]{$2$}}
          \put(-5,60){\makebox(0,0)[r]{$3$}}
          \put(38,63){\makebox(0,0)[l]{$4$}}
          \put(65,65){\makebox(0,0)[l]{$6$}}
          \put(96,62){\makebox(0,0)[l]{$8$}}
          \put(125,60){\makebox(0,0)[l]{$10$}}
          \put(60,95){\makebox(0,0)[b]{$1$}}

          \put(60,0){\circle*{3}}
          \put(0,30){\circle*{3}}
          \put(30,30){\circle*{3}}
          \put(60,30){\circle*{3}}
          \put(90,30){\circle*{3}}
          \put(120,30){\circle*{3}}
          \put(0,60){\circle*{3}}
          \put(30,60){\circle*{3}}
          \put(60,60){\circle*{3}}
          \put(90,60){\circle*{3}}
          \put(120,60){\circle*{3}}
          \put(60,90){\circle*{3}}
        \end{picture}
      } 

    \end{picture}
    \caption{Join semilattice induced by the OMIA of Table~\ref{tab:chcounterex}(i).
     When
     combined with the partial functions $a^\perp_b$ of Table~\ref{tab:chcounterex}(ii),
     it provides an example of an orthomodular join semilattice (OJS) that is not
     a semi-orthomodular lattice (SOL).
\label{fig:chcounterex}}
\end{figure}
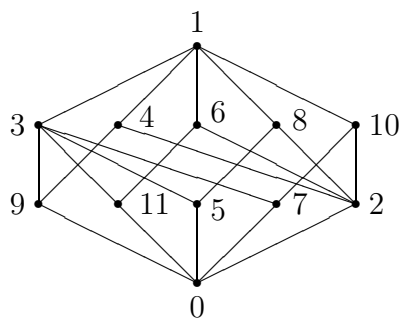

Theorem~\ref{th:omia-ne-oia} tells us that the axioms of OIA cannot be derived from the
axioms of OMIA.  In particular, this proves that the axioms of OMIA are incomplete.
In other words there exist equational theorems of OML, expressible purely in terms of the
Dishkant implication, that cannot be proved from the axioms of OMIA.  Axiom OI3 of OIA
is one such example.  Another example that does not hold in all OMIAs is
the ``implication version of the orthomodular law'' of
\cite{abbott76}:
\begin{eqnarray}
a\le b\le c & \mbox{implies} & c=(ca)b.\label{impoml}  
\end{eqnarray}
The OMIA of Table~\ref{tab:chcounterex}(i) violates this law as can be seen
by choosing $a=0$, $b=2$, $c=4$.

Similarly, not all OJSs are SOLs.  The join semilattice of
Figure~\ref{fig:chcounterex} along with the $a^\perp_b$ operations
specified by Table~\ref{tab:chcounterex}(ii) define an OJS.  However,
this OJS violates condition C of Definition~\ref{D:sol}, as can be seen
by choosing $a=0$, $b=2$, $c=4$.  [Although this example also happens to
be a lattice, we remind the reader that in general join semilattices are
not bounded below.]

In conclusion, we have shown that the axioms of OMIA are not complete,
since in particular they are strictly weaker than the axioms of OIA.  On
the other hand, the completeness of the axioms for OIA is apparently not
known \cite{harde81a}.  Future work towards seeking a complete Dishkant
implicational fragment of OML theory might prove more fruitful by
investigating OIA, rather than OMIA, as a starting point.

\section{\large Implication algebra based on the Sasaki implication}
\label{sec:sasaki}

Apparently the only other pure implicational fragment of OML theory
that has been studied are ``quasi-implicational algebras'' based on
the Sasaki implication $\to_1$ \cite{harde81a,harde81b}.

\begin{definition}\label{D:qsia} \cite{harde81a} A quasi-implication
algebra ({\rm QSIA}) is an algebraic structure $\langle{\mathcal A},\circ
\rangle$ with a single binary operation that satisfies:

\smallskip
{\bf QS1}\quad$(a\circ b)\circ a=a$

\smallskip
{\bf QS2}\quad$(a\circ b)\circ (a\circ c)=(b\circ a)\circ (b\circ c)$

\smallskip
{\bf QS3}\quad$((a\circ b)\circ(b\circ a))\circ a=((b\circ a)\circ
     (a\circ b))\circ b$

\end{definition}

QSIA is sound for the Sasaki implication in the
sense that if the binary operation $\circ$ is replaced throughout by
$\rightarrow _1$, each axiom becomes an equation that holds in all OMLs.

An important result is that QSIA is also {\em complete} in the sense
that when $\circ$ is interpreted as $\rightarrow _1$, its theorems are
precisely those equational theorems of OML theory where each side of an
equation is expressible purely in terms of polynomials built from
$\rightarrow _1$ \cite{harde81b}.

A simple observation also shows that every QSIA induces an OIA (and
an OMIA by Theorem~\ref{th:oia-omia}).

\begin{theorem} Every {\rm QSIA} induces an {\rm OIA} under the definition
$ab \>=^{\rm def}\> (b\circ a)\circ(a\circ b)$.
\end{theorem}
\begin{proof}
In any OML, $a\to_2 b=(b\to_1 a)\to_1 (a\to_1 b)$.  Since OIA is
sound for $\to_2$ in OML, we can replace $\to_2$ for $\cdot$ throughout
the axioms of OIA, then express them in terms of $\to_1$ per this
equation, to obtain equations built from $\to_1$ that hold in
all OMLs.  By the completeness of QSIA, each of these equations
is provable under QSIA after substituting $\circ$ for $\to_1$.
\end{proof}

The converse, that every OIA induces a QSIA, is not obtainable with
a simple substitutional definition since it is impossible to express
$\to_1$ in terms of a polynomial built from $\to_2$.  Thus there
is a sense in which QSIA is ``richer'' than OIA.  Whether there exists
a more indirect isomorphism between OIA and QSIA is unknown.

\section{\large The relationships among the various implications}
\label{sec:relation}

  From the observation in the previous section that $\to_2$ can be
expressed in terms of $\to_1$, we were led to investigate the other
ways of expressing one implication in terms of another.

\begin{table}[hbtp]
{\begin{center}
\begin{tabular}{c|l}
$a\to_i b$ & \multicolumn{1}{|c}{$a\to_i b$ expressed in terms of other implications} \\
        \hline
$a\to_0 b=$ &   $((b\to_1 a)\to_1 a)\to_1 b$,\quad $a\to_3 (a\to_3 b)$,\\
              & \qquad    $((a\to_4 b)\to_4 b)\to_4 b$,\quad $(b\to_5 a)\to_5 (a\to_5 b)$,\\
              & \qquad      $a\to_5 ((b\to_5 a)\to_5 b)$  \\
$a\to_1 b=$ &   $a\to_5 (a\to_5 b)$ \\
$a\to_2 b=$ &   $(b\to_1 a)\to_1 (a\to_1 b)$,\quad $(b\to_3 a)\to_3 (a\to_3 b)$,\\
              & \qquad
                  $((a\to_3 b)\to_3 a)\to_3 b$,\quad $a\to_4 (a\to_4 b)$,\\
              & \qquad $((a\to_5 b)\to_5 b)\to_5 b$,\quad $((b\to_5 a)\to_5 a)\to_5 b$,\\
              & \qquad    $((a\to_5 b)\to_5 a)\to_5 b$ \\
$a\to_3 b=$ &   $(a\to_1 (b\to_1 a))\to_1 ((b\to_1 a)\to_1 (a\to_1 b))$,\\
              & \qquad    $(a\to_5 (b\to_5 a))\to_5 (a\to_5 b)$ \\
$a\to_4 b=$ &   $((b\to_1 a)\to_1 a)\to_1 (a\to_1 b)$,\\
              & \qquad    $(((b\to_5 a)\to_5 b)\to_5 b)\to_5 (a\to_5 b)$ \\
$a\to_5 b=$ &   [none other than $a\to_5 b$ itself]
\end{tabular}
\end{center}}
    \caption{The shortest expressions of the implications in terms of others.
      (When there are more than one shortest, all are shown.)
\label{tab:impgen}}
\end{table}

With the assistance of the computer programs {\tt beran.c} and {\tt
bercomb.c} (obtainable from the authors), we exhausted the possibilities
and obtained the results in Table~\ref{tab:impgen}, where we show
shortest expressions for each implication that can express other ones.
For completeness we also include the classical implication $\to_0$.

\begin{table}[hbtp]
{\begin{center}
\begin{tabular}{c|l}
$\to_i$ & \multicolumn{1}{|c}{Beran numbers for $\to_i$ polynomials
              with two generators} \\ \hline
$\to_0$ & 22 28 39 44 93 94 96 \\
$\to_1$ & 22 23 28 29 30 32 38 39 44 45 46 48 54 60 61 62 64 71 76 77 78 80 \\
    & \qquad 78 80 86 87 92 93 94 96 \\
$\to_2$ & 22 29 39 46 92 96 \\
$\to_3$ & 22 23 28 29 30 32 38 39 44 45 46 48 86 87 92 93 94 96 \\
$\to_4$ & 22 28 29 32 39 44 46 48 55 60 62 64 70 76 77 80 86 87 92 93 94 96 \\
$\to_5$ &  6  7 12 13 14 16 22 23 28 29 30 32 38 39 44 45 46 48 54 55 60 61 62 \\
   &  \qquad 64 70 71 76 77 78 80 86 87 92 93 94 96
\end{tabular}
\end{center}}
    \caption{The Beran numbers for all possible polynomials with two
generators  built from implications $\to_i$.
\label{tab:berangen}}
\end{table}

Any OML polynomial with two generators (variables) corresponds to one of
96 possible expressions (Beran expressions).  For brevity, we label
Beran expressions with the numbers assigned in \cite[p.~82]{beran}.  The
Beran numbers for implications $a\to_i b$ are 94, 78, 46, 30, 62, and 14
for $i=0,\ldots,5$ respectively.  We refer the reader to
\cite[p.~82]{beran} for the expressions corresponding to any Beran
numbers we do not show explicitly.

Polynomials built from the $\to_2$ operation generate only 6 of the 96
possible expressions:  $a$ (with Beran number 22), $b\to_2 a$ (29), $b$
(39), $a\to_2 b$ (46), $a \cup b$ (92), and $1$ (96).

The other quantum implications $\to_1$, $\to_3$, $\to_4$, and $\to_5$
generate respectively 28, 18, 22, and 36 Beran expressions.  In
Table~\ref{tab:berangen} we show their Beran numbers.  In particular, we
note from this table that the intersection of the sets of Beran numbers
for all quantum implications is the same as the set of Beran numbers for
$\to_2$, and the union of them is the same as the set of Beran numbers
for $\to_5$.

Thus $\to_5$ is the ``richest'' and $\to_2$ the ``poorest'' generator.
In particular, $\to_5$ can generate all other implications, and all
quantum implications can generate $\to_2$.

\section{\large Quantum implication algebra}
\label{sec:unified}

In \cite{mpijtp98} we showed that a single, structurally identical
expression, that holds when its operation is any one of quantum
implications, can represent the join operation:
\begin{eqnarray}
a\cup b&=&(a\rightarrow_i b)\rightarrow_i(((a
\rightarrow_i b)\rightarrow_i(b \to_i a))\rightarrow_i a)
\end{eqnarray}
holds in any OML for $i=1,\ldots,5$.  This observation allowed us to
construct, by adding a constant $0$, an OML-equivalent
algebra with an (unspecified) quantum implication as its only
binary operation.
Prompted by this result, we investigated the possibility of a purely
implicational system having a single unspecified quantum implication as its
sole operation.

In the previous section we observed that the $\to_2$ implication is
unique in that it can be generated by any one of the other quantum
implications.  It turns out that there exists a {\em single} expression
with an operation which, if replaced throughout by any one of the
quantum implications $\to_i, i=1,\ldots,5$, will evaluate to $\to_2$.

\begin{theorem}\label{th:unif2}The equation
\begin{eqnarray}
a\to_2 b&=&(b\to_i (b\to_i a))\to_i (((a\to_i b)\to_i a)\to_i b)
\end{eqnarray}
holds in any {\rm OML}, for all $i\in\{1,2,3,4,5\}$.
\end{theorem}
\begin{proof}The verification is straightforward.
\end{proof}

This allows us to define an implicational algebra that works when the
binary operation is interpreted as any quantum implication.

\begin{definition}\label{def:star} A quantum implication algebra
({\rm QIA}) is an algebraic structure
$\langle{\mathcal A},\bullet \rangle$ with a single
binary operation that satisfies:

\smallskip
{\bf Q1}\quad$(a\star b)\star a=a$

\smallskip
{\bf Q2}\quad$(a\star b)\star b=(b\star a)\star a$

\smallskip
{\bf Q3}\quad$a\star ((b\star a)\star c)=a\star c$

\parindent=0pt
\smallskip
where $a\star b {\buildrel\rm def\over=} (b\bullet (b\bullet a))\bullet
   (((a\bullet b)\bullet a)\bullet b)$
\end{definition}

\begin{theorem}
{\em QIA} is sound for any quantum implication $\to_i, i=1,\ldots,5$ in the
sense that if the binary operation $\bullet$ is replaced throughout by
$\to_i$, each axiom becomes an equation that holds in all {\rm OML}s.
\end{theorem}
\begin{proof}The axioms of QIA are the same as the axioms of OIA with
$\star$ substituted for $\cdot$.  Soundness follows from
Theorem~\ref{th:unif2} and the soundness of OIA.
\end{proof}

\begin{theorem}\label{th:unifind}Every {\rm QIA} induces an {\rm OIA} under the definition
$ab  \>=^{\rm def}\> a\star b$.
\end{theorem}
\begin{proof}The axioms of QIA become the axioms of OIA when
$\cdot$ is substituted for $\star$.
\end{proof}

As a corollary, every {\rm QIA} induces a semi-orthomodular lattice
(SOL), following the proof of \cite{abbott76}.  Conversely, every SOL
induces a QIA by Theorem~\ref{th:qsol}(ii) below.

\begin{lemma}\label{lem:q2}The following equation holds in every {\rm OIA}
   (and every {\rm OMIA}):
\begin{eqnarray}
ab&=&(b(ba))(((ab)a)b)\label{eq:q2}
\end{eqnarray}
\end{lemma}
\begin{proof}
We show this equation holds in OMIA, and that it holds in OIA follows from
Theorem~\ref{th:oia-omia}.  (i) $b(ba)=((ba)b)(ba)=ba$ using O3 twice.
(ii) $((ab)a)b=ab$ using O3.
(iii) $ab\le(ba)(ab)$ using O2.  (iv) $a\le ba$ using O2, so $(ba)(ab)\le a(ab)=ab$ by
Lemma 4(ix) of \cite{langer1} and O3.  (v) From (iii) and (iv), we
have $ab=(ba)(ab)$ by
Lemma 4(vi) of \cite{langer1}.  Substituting (i) and (ii) into this we obtain
the result.
\end{proof}

\begin{theorem}\label{th:qsol}
(i) Every {\rm OIA} induces a {\em QIA} under the definition
$a\bullet b \>=^{\rm def}\> ab$.
(ii) Every {\rm SOL} induces a {\em QIA} under the
definition $a\bullet b \>=^{\rm def}\> (a \cup b)^\perp_b$.
\end{theorem}
\begin{proof}
(i) We convert each axiom of OIA by simultaneously expanding each
occurrence of $\cdot$ into the right-hand side of Eq.~\ref{eq:q2}.
Substituting $\bullet$ for $\cdot$ throughout, we obtain the axioms of
QIA.  (ii) Immediate from (i) and Theorem~\ref{th:oia-sol}(ii).
\end{proof}

The system QIA that we have given is not complete.  For example, the
equation $a\bullet (a\bullet a)=a\bullet a$ is not a theorem of QIA
(by virtue of the structure of Axioms Q1---Q3) even
though it is sound for all quantum implications.  QIA was devised for our
purposes to be sufficient to induce an OIA, and nothing more.  What such
a complete axiomatisation would look like, and even whether it can be
finitely axiomatised, remain open problems.

\section{\large Unified quantum implication algebras}
\label{sec:minimal}

In the previous section we have shown how one can construct an implication
algebra with the same axioms for all five possible implications.  If we
are interested in specific implications, we can construct more
specialised algebras with somewhat shorter axioms if we---in
Def.~\ref{def:star}---chose $a\star b {\buildrel\rm def\over=}a\bullet
b$ (for $\to_2$), or $(b\bullet a)\bullet (a\bullet b)$ (for $\to_1$ and
$\to_3$), or $((a\bullet b)\bullet a)\bullet b$ (for $\to_3$ and
$\to_5$), or $a\bullet(a\bullet b)$ (for $\to_4$).  Another possible
choice is $a\star b {\buildrel\rm def\over=} (a\sqcup b)\bullet b$ where
$a\sqcup b$ is defined as in Def.~\ref{def:mqia}.  None of these
algebras is proven to be complete (and therefore ``maximal'') in the
sense of QSIA (see Section \ref{sec:sasaki}).

On the other hand, one can take a more direct approach of
finding implication algebras which would comply with the
following objectives:
\begin{enumerate}
\item proving that the algebras are partially ordered sets bounded
from above;
\item  proving that the algebras induce join semilattices in which 
every principal order filter generates an orthomodular lattice;
\item proving that the algebras, when they contain a smallest element
0, can induce orthomodular lattices.
\end{enumerate}

While QIA satisfies these objectives, its axioms are very long.  Systems
designed specifically with these objectives as their goal can have
shorter axioms that are easier to work with.  Here we give examples
of such systems.

\begin{definition}\label{def:mqia}  A unified quantum implication
algebras {\rm UQIAi} are algebraic structures
$\langle{\mathcal A},\bullet\rangle$ with single binary operations
that satisfy:

\smallskip
{\bf UQ1}\quad$a\bullet a=b\bullet b$

\smallskip
{\bf UQ2}\quad $a\bullet (a\sqcup b)=1$

\smallskip
{\bf UQ3}\quad $b\bullet (a\sqcup b)=1$

\smallskip
{\bf UQ4}\quad $a\bullet 1=1$

\smallskip
{\bf UQ5}\quad $a\bullet b=1\quad \&\quad b\bullet a=1\quad\Leftrightarrow\quad a=b$

\smallskip
{\bf UQ6}\quad $a\bullet b=1\quad \&\quad b\bullet c=1\quad\Rightarrow\quad a\bullet c=1$

\smallskip
{\bf UQ7}\quad $a\bullet c=1\quad \&\quad b\bullet c=1\quad\Rightarrow\quad
(a\sqcup b)\bullet c=1$

\smallskip
{\bf UQ8}\quad $b\bullet a=1\quad\Rightarrow\quad
a\sqcup (a\bullet b)=1$

\smallskip
{\bf UQ9}\quad $b\bullet a=1\quad\Rightarrow\quad ((a\bullet b)\bullet b)\bullet a=1$

\smallskip
{\bf UQ10}\quad $b\bullet a=1\quad\Rightarrow\quad a\bullet ((a\bullet b)\bullet b)=1$

\smallskip
{\bf UQ11}\quad $b\bullet a=1\quad \&\quad c\bullet a=1\quad \&\quad c\bullet b=1\quad
\Rightarrow\quad (a\bullet c)\bullet (b\bullet c)=1$

\smallskip
{\bf UQ12}\quad  $c\bullet a=1\quad \&\quad c\bullet b=1\quad \&\quad a\bullet b=1
\quad \&\quad a\sqcup(b\bullet c)=1$

$\qquad\qquad\quad\Rightarrow\quad b\bullet a=1$

\medskip
\noindent
where $1{\buildrel\rm def\over\Leftrightarrow}a\bullet a$ and $a\sqcup
b$ means either $(a\bullet b)\bullet b$ (for either $\to_2$ or $\to_5$),
or $((a\bullet b)\bullet (b\bullet a))\bullet a$ (for either $\to_1$ or
$\to_3$), or $(a\bullet (a\bullet b))\bullet b$ (for $\to_4$), or
$((((a\bullet b)\bullet (b\bullet a))\bullet a)\bullet b)\bullet b$ (for
$\to_i$, $i=1,\dots,5$).
\end{definition}

The above non-unique ways of expressing  $a\sqcup b$ is a consequence
of the fact that in an OML one cannot express $a\sqcup b$ in unique
ways by using nothing but implications.
(By ``unique'' we mean that an expression, in an OML, evaluates to
$a\cup b$ for only one of the five implications and no others.)
In an OML one can use
implications \it and\/ \rm complements in, e.g., the following way:
\begin{enumerate}
\smallskip
\item\qquad $a\cup b=b^\perp\to_1((b^\perp\to_1a^\perp)^\perp\to_1
(b\to_1a^\perp)^\perp)^\perp$
\smallskip
\item\qquad $a\cup b=(b^\perp\to_2(b\to_2(b^\perp\to_2a^\perp)^\perp
)^\perp)^\perp\to_2a$
\smallskip
\item\qquad $a\cup b=b^\perp\to_3(b^\perp\to_3a)$
\smallskip
\item\qquad $a\cup b=a^\perp\to_4(b^\perp\to_4a)$
\smallskip
\item\qquad $a\cup b=(a\to_5b^\perp)\to_5(b^\perp\to_5a)$
\end{enumerate}
\noindent
Here, e.g., no one of $\to_i,\ i=1,\dots,5$ except $\to_3$ would satisfy the
3rd line. However, one can again express implications by each other, so
that, in the end, ambiguous expressions are equally proper as these ones.

Like QIA, algebras UQIA(i) are fragments of ``maximal'' algebras for
their respective implications or sets of implications.
However, they are sufficiently strong to accomplish our
objectives above.  Among other possibilities, they could be useful
starting points in a search for maximal algebras (which are currently
open problems for all cases except the $\to_1$ of QSIA).

\begin{theorem}\label{th:part-ord}
Every unified quantum implication algebra
{\rm UQIA}=$<{\mathcal A},\bullet>$
determines an associated partially ordered set with an upper
bound under:
\begin{eqnarray}
a\le b\quad{\buildrel\rm def\over\Leftrightarrow}\quad ab=1
\label{eq:order}
\end{eqnarray}
\end{theorem}

\begin{proof}
We have to prove

(1) $a \le a$

(2) $a\le b\quad \&\quad b\le a\quad\Rightarrow\quad a=b$

(3) $a\le b\quad \&\quad b\le c\quad\Rightarrow\quad a\le c$

(4) $a\le 1$

\noindent
(1) follows from the definition of 1 and Eq.~(\ref{eq:order}).

\noindent
(2) follows from {\bf UQ5} and  Eq.~(\ref{eq:order}).

\noindent
(3) follows from {\bf UQ6} and  Eq.~(\ref{eq:order}).

\noindent
(4) follows from {\bf UQ4} and  Eq.~(\ref{eq:order}).

\end{proof}

\begin{theorem}\label{th:join-semi}
$<{\mathcal A},\le,\cup,1>$ in which one defines:
$a\cup b\ {\buildrel\rm def\over\Leftrightarrow}\ a\sqcup b$,
is a join semilattice.
\end{theorem}

\begin{proof}
We have to prove that $a\cup b=\ ${\rm sup}$\{a,b\}$, i.e., that
the following conditions are satisfied:

(1) $a\le a\cup b$

(2) $b\le a\cup b$

(3) $a\le c\quad \&\quad b\le c\quad\Rightarrow\quad
a\cup b\le c$

\noindent
(1) follows from {\bf UQ2}

\noindent
(2) follows from {\bf UQ3}

\noindent
(3) follows from {\bf UQ7}
\end{proof}

\begin{theorem}\label{th:semi-oml}
If $m\in {\mathcal A}$ is a fixed element and one defines:
\begin{eqnarray}
a^\perp_m\quad{\buildrel\rm def\over\Leftrightarrow}\quad am,
\label{eq:neg}
\end{eqnarray}
and
\begin{eqnarray}
a\cap b \quad{\buildrel\rm def\over\Leftrightarrow}\quad ((am)\cup(bm))m
\ \qquad for\ \ a,b\in{\mathcal I}_m\label{eq:meet}
\end{eqnarray}
then $<{\mathcal I}_m,\cup,\cap,m,1,a^\perp_m>$, where
${\mathcal I}_m=\{a\in {\mathcal A}\>|\>m\le a\}$ is the principal
order filter generated by $m$, is an orthomodular lattice.
\end{theorem}

\begin{proof}
We have to prove that the following conditions
for the above ($m\le a$) are satisfied:
\begin{eqnarray}
&(1)& \quad a\cup a^\perp_m=1\label{eq:ortho-1}\\
&(2)&\quad  a^{\perp\perp}_{mm}=a\label{eq:ortho}\\
&(3)&\quad  a\le b\quad\Rightarrow\quad b^\perp_m\le
a^\perp_m \label{eq:ortho-lr}
\end{eqnarray}

\noindent
(1) follows from {\bf UQ8} since $ma=1$ holds for any $a$.

\noindent
(2) follows from {\bf UQ9}, {\bf UQ10}, and {\bf UQ5}.

\noindent
(3) follows from {\bf UQ11} by taking $c=m$ since $ma=1$ and
$mb=1$ hold for any $a$ and $b$.

\smallskip
Then we have to prove that $a\cap b=\ ${\rm inf}$\{a,b\}$,
i.e., that the following conditions are satisfied:

(1) $a\cap b\le a$

(2) $a\cap b\le b$

(3) $a\le b\quad \&\quad a\le c\quad\Rightarrow\quad
a\le b\cap c$

\noindent
(1) follows from {\bf UQ2} and Eq.~(\ref{eq:ortho-lr}).

\noindent
(2) follows from {\bf UQ3} and Eq.~(\ref{eq:ortho-lr}).

\noindent
(3) follows from {\bf UQ7}  and Eqs.~(\ref{eq:ortho-lr})
and (\ref{eq:ortho}).

In the end we have to prove the orthomodularity. By taking
$c=m$, we get $ma=1$ and $mb=1$, i.e., $m\le a$ and $m\le b$
for any $a$ and $b$ so that {\bf UQ12} gives us the
orthomodularity:
$$a\le b\quad \&\quad a\cup b^\perp_m=1\quad\Rightarrow\quad b\le a$$
\end{proof}

\begin{corollary}\label{th:isom} A {\rm UQIA} with a smallest element
$0$, i.e.\ satisfying the axiom $0\bullet a=1$, induces an {\rm OML} under the
definitions $a\cup b {\buildrel\rm def\over=} a\sqcup b$ and $a'
{\buildrel\rm def\over=} a\bullet 0$.  A {\rm QIA} with a smallest element 0
induces an {\rm OML} under the definitions $a\cup b {\buildrel\rm
def\over=} (a\star b)\star b$ and $a' {\buildrel\rm def\over=} a\star
0$.
\end{corollary}

\begin{proof}Straightforward.\end{proof}

\section{\large Conclusion}
\label{sec:conclusion}

We have investigated implication algebras for orthomodular lattices.
We have first compared the systems previously given by \cite{abbott76}
(OIA, orthoimplication algebra), \cite{langer1} (OMIA, orthomodular
implication algebra), and \cite{harde81a,harde81b} (QSIA,
quasi-implication algebra).

In Sec.~\ref{sec:rel} we proved that the axioms of OMIA can be derived
from the axioms of OIA but not vice-versa. In other words, we have shown
that the axioms of OMIA are not complete. In particular, the implication
version of the orthomodular law does not hold in OMIA contrary to its
name (\it orthomodular\/ \rm implication algebra). Whether OIA is complete
in the sense of Hardegree's QSIA remains an open problem. For, QSIA's
theorems are precisely those equational theorems of the OML theory where
each side of an equation is expressible purely in terms of polynomials
built from the corresponding OML (Sasaki) implication. If one wanted
to attack the completeness problem along the way taken by Hardegree,
we conjecture that the relevance implication ($i=5$) would be the
most promising with respect to Table \ref{tab:impgen}. Also, we would
like to point out that the first axiom of both OIA and QSIA
is the OML property $a\cup b=b\cup a$ expressed by means of implications.
Their second axiom is the OML property $a=a$, where the left $a$
is given as its shortest implication presentation involving two
variables. \cite{mpqo02}

In Sec.~\ref{sec:relation} we investigate the other
ways of expressing one implication in terms of another and in
Sec.~\ref{sec:unified} we combined the obtained results to show how
one can formulate quantum implication
algebras, QIA's which keep the same form for all five possible
implications from OML thus capturing an essential
property that is common to all quantum implications.

In Sec.~\ref{sec:minimal} we formulated unified quantum
implication algebras (UQIA's) for all implications. They are so week
that they do not yield a single axiom of either OIA or QSIA.
Still, their join semilattices with 0 induce orthomodular lattices.

An open problem is devising a maximal extensions of QIA and UQIA that
are complete, in the sense that its theorems are precisely those equational
theorems of OML theory that hold regardless of which quantum implication $\to_i,
i=1,\ldots,5$ we substitute for $\bullet$.  A complete axiomatisation of
QIA and UQIA would be interesting because it would provide a general way to
explore properties that are common to all quantum implications.  It
would also provide a way around philosophical debates about which
quantum implication is the ``proper'' or ``true'' implication for
quantum logic, since any of its results immediately apply to whichever
one we prefer.  And, finally, it might reduce concerns about being led
astray by ``toy'' systems \cite{urqu} since we would not be focusing on
the specialised properties of any one implication in particular.

\vfill\eject

\bigskip\bigskip
\parindent=0pt
{\large\bf ACKNOWLEDGMENTS}

\parindent=20pt
\bigskip

M.~P.~acknowledges supports of the Ministry of Science of Croatia
through the project 0008222.

The authors would like to thank to William McCune, Argonne National Lab,
Argonne IL, U.~S.~A., for his assistance in using the program Mace4
({\tt http://www.mcs.anl.gov/\~{}mccune/mace4/}), which found the initial
version of Table~\ref{tab:chcounterex}.

\vfill\eject

\medskip

\end{document}